\documentclass[journal=nalefd,manuscript=letter,layout=traditional]{achemso}

\setkeys{acs}{etalmode=truncate}
\setkeys{acs}{maxauthors=0}
\setkeys{acs}{articletitle=true}

\usepackage{amssymb}

\usepackage{graphicx}% Needed to include png/jpg/pdf figure files
\usepackage{amsmath}
\usepackage{bm}
\usepackage{color}
\usepackage[normalem]{ulem}
\usepackage{siunitx}
\usepackage{xr}
\usepackage{braket}
\title{Intervalley Scattering of Interlayer Excitons in a MoS$_2$/MoSe$_2$/MoS$_2$ Heterostructure in High Magnetic Field}

\author{Alessandro Surrente}
\affiliation{Laboratoire National des Champs Magn\'etiques Intenses, UPR 3228, CNRS-UGA-UPS-INSA, 38042 Grenoble and 31400 Toulouse, France}
\author{{\L}ukasz\ K{\l}opotowski}
\affiliation{Institute of Physics, Polish Academy of Sciences, Al.\ Lotnik{\'o}w 32/46, 02-668 Warsaw, Poland}
\author{Nan Zhang}
\affiliation{Laboratoire National des Champs Magn\'etiques Intenses, UPR 3228, CNRS-UGA-UPS-INSA, 38042 Grenoble and 31400 Toulouse, France}
\author{Michal Baranowski}
\affiliation{Laboratoire National des Champs Magn\'etiques Intenses, UPR 3228, CNRS-UGA-UPS-INSA, 38042 Grenoble and 31400 Toulouse, France}
\alsoaffiliation{Department of Experimental Physics, Faculty of Fundamental Problems of Technology, Wroclaw University of Science and Technology, 50-370 Wroclaw, Poland}
\author{Anatolie A.\ Mitioglu}
\affiliation{High Field Magnet Laboratory (HFML -- EMFL), Radboud University, 6525 ED Nijmegen, The Netherlands}
\author{Mariana V.\ Ballottin}
\affiliation{High Field Magnet Laboratory (HFML -- EMFL), Radboud University, 6525 ED Nijmegen, The Netherlands}
\author{Peter C.\ M.\ Christianen}
\affiliation{High Field Magnet Laboratory (HFML -- EMFL), Radboud University, 6525 ED Nijmegen, The Netherlands}
\author{Dumitru Dumcenco}
\affiliation{Electrical Engineering Institute and Institute of Materials Science and Engineering, \'{E}cole Polytechnique F\'{e}d\'{e}rale de Lausanne, CH-1015 Lausanne, Switzerland}
\author{Yen-Cheng Kung}
\affiliation{Electrical Engineering Institute and Institute of Materials Science and Engineering, \'{E}cole Polytechnique F\'{e}d\'{e}rale de Lausanne, CH-1015 Lausanne, Switzerland}
\author{Duncan K.\ Maude}
\affiliation{Laboratoire National des Champs Magn\'etiques Intenses, UPR 3228, CNRS-UGA-UPS-INSA, 38042 Grenoble and 31400 Toulouse, France}
\author{Andras Kis}
\affiliation{Electrical Engineering Institute and Institute of Materials Science and Engineering, \'{E}cole Polytechnique F\'{e}d\'{e}rale de Lausanne, CH-1015 Lausanne, Switzerland}
\author{Paulina Plochocka}\email{paulina.plochocka@lncmi.cnrs.fr}
\affiliation{Laboratoire National des Champs Magn\'etiques Intenses, UPR 3228, CNRS-UGA-UPS-INSA, 38042 Grenoble and 31400 Toulouse, France}

\keywords{Transition metal dichalcogenides, van der Waals heterostructures, interlayer exciton, magnetophotoluminescence, valley polarization}

\begin{document}

%\date{\today}

\begin{abstract}

Degenerate extrema in the energy dispersion of charge carriers in solids, also referred to as valleys, can be regarded as a
binary quantum degree of freedom, which can potentially be used to implement valleytronic concepts in van der Waals
heterostructures based on transition metal dichalcogenides. Using magneto-photoluminescence spectroscopy, we achieve a deeper
insight into the valley polarization and depolarization mechanisms of interlayer excitons formed across a
MoS$_2$/MoSe$_2$/MoS$_2$ heterostructure. We account for the non-trivial behavior of the valley polarization as a function of the magnetic field by
considering the interplay between exchange interaction and phonon mediated intervalley scattering in a system consisting of
Zeeman-split energy levels. Our results represent a crucial step towards the understanding of the properties of interlayer
excitons, with strong implications for the implementation of atomically thin valleytronic devices.
\end{abstract}

\maketitle

In close analogy with spin and its use in quantum information processing \cite{awschalom2013quantum}, the valley pseudospin
\cite{xu2014spin} can potentially be used to encode, store and transfer information. While valley physics has been investigated
in a variety of materials, including AlAs \cite{shkolnikov2002valley}, silicon \cite{salfi2014spatially}, diamond
\cite{isberg2013generation}, bismuth \cite{zhu2012field}, and graphene \cite{gorbachev2014detecting}, the lack of a direct band
gap in these systems precludes optically addressing and reading out the valley degree of freedom. In monolayer transition metal
dichalcogenides (TMDs), the valley degree of freedom corresponds to a direct band gap in the visible range at the nonequivalent
but degenerate K$^+$ and K$^-$ points of the Brillouin zone. This, along with the locking of the the spin and
valley degrees of freedom \cite{xiao2012coupled}, allows to optically initialize \cite{mak2012control}, manipulate
\cite{ye2017optical} and read out \cite{hao2016direct} the valley pseudospin using circularly polarized light. The large exciton
binding energy in these materials \cite{chernikov2014exciton} is reflected in a very large dipole moment, which induces sub-ps
radiative lifetimes \cite{moody2015intrinsic}. This, combined with very efficient intervalley scattering \cite{zhu2014exciton}
mediated via electron-hole exchange interaction \cite{baranowski2017dark}, strongly limits their usefulness of monolayer TMDs in practical
valleytronic devices for quantum information science.

A possibility to overcome these limitations is offered by van der Waals heterostructures, obtained by vertically stacking
monolayers of different TMDs. Photoexcited charge carriers are quickly separated \cite{hong2014ultrafast} due to the type II
band alignment exhibited by heterobilayers \cite{kang2013band}, forming a quasiparticle referred to as interlayer exciton. The
spatial separation of the charges leads to a significant increase (up to five order of magnitudes, $\sim\SI{100}{\nano\s}$) of
the recombination life time of interlayer excitons \cite{miller2017long,baranowski2017probing}. Moreover, in analogy to intralayer
excitons, valley polarization of interlayer excitons can be injected via optical means. The reduced spatial overlap of the
electron-hole wavefunctions leads to a dramatically decreased electron-hole exchange interaction, which results in long-lived
valley polarization (up to \SI{40}{\nano\s}) \cite{rivera2016valley}. These properties make interlayer excitons, formed across
van der Waals heterostructures, ideally suited for valleytronic applications.

The manipulation of the valley degree of freedom via the application of magnetic field has been successfully demonstrated for
intralayer excitons \cite{li2014valley,aivazian2015magnetic,srivastava2015valley,macneill2015breaking,Mitioglu15}. This approach
has recently been extended to interlayer excitons, where the observation of a giant valley Zeeman splitting and a subsequent
near-unity valley polarization has been enabled by the \SI{60}{\degree} tilt angle between mechanically exfoliated
MoSe$_2$/WSe$_2$ monolayers \cite{nagler2017giant}.  However, a fundamental understanding of the dynamics of the interlayer
exciton population as well as valley depolarization mechanism is still lacking.

Here, we achieve a deeper insight into the properties of interlayer excitons by performing detailed magneto-photoluminescence
(magnetoPL) spectroscopy of a heterostructure formed by MoS$_2$ and MoSe$_2$ monolayers \cite{baranowski2017probing,
surrente2017defect}. The materials which compose our heterostructures are lattice mismatched. This leads to the formation of a moir{\'e} pattern \cite{zhang2017interlayer,pan2018quantum}, which has been shown to influence the electronic and valley properties of interlayer excitons \cite{kang2013electronic,yu2017moire,wu2018theory}. We lift the valley degeneracy by applying magnetic fields up to \SI{28}{\tesla} in the Faraday
configuration. We observe a population imbalance in the Zeeman split valleys, which allows us to precisely control the valley polarization from 0 to almost 100\% by applying the magnetic field. For the first time, we describe the magnetic field dependence of the valley polarization of a van der Waals heterostructure by a model, which accounts for the observed intervalley relaxation via an interplay between exchange and phonon driven intervalley scattering in Zeeman split levels.

\begin{figure*}[h!]
\centering
\includegraphics[width=1.0\linewidth]{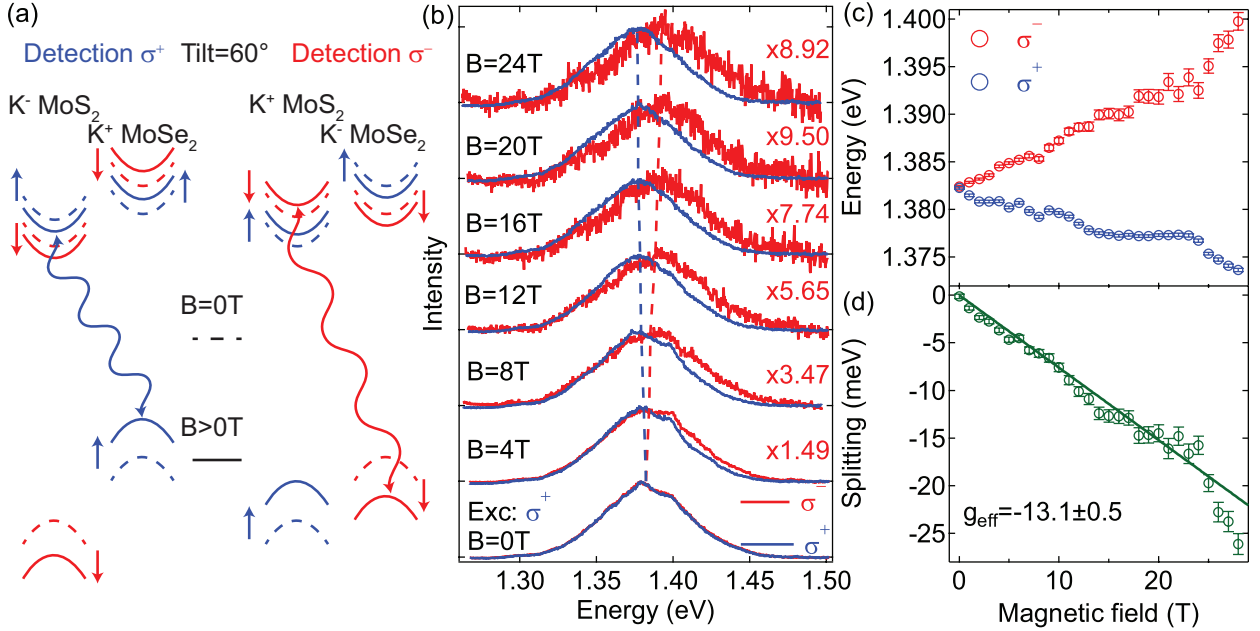}
\caption{(a) Configuration of the relevant band edges at zero and high magnetic field, assuming a \SI{60}{\degree} stacking angle
between the monolayers. The spin of the bands is color-coded and indicated by an arrow of the same color. Wavy arrows depict
dipole-allowed optical transitions. (b) Interlayer exciton magnetoPL spectra up to \SI{24}{\tesla}. The spectra detected with
$\sigma^-$ polarization have been rescaled. The dashed lines are a guide for the eye. (c)
Transition energies, and (d) energy splitting of interlayer exciton as a function of magnetic field.}
\label{fig:MagnetoPLenergies}
\end{figure*}

\begin{figure*}[h!]
\includegraphics[width=1.0\linewidth]{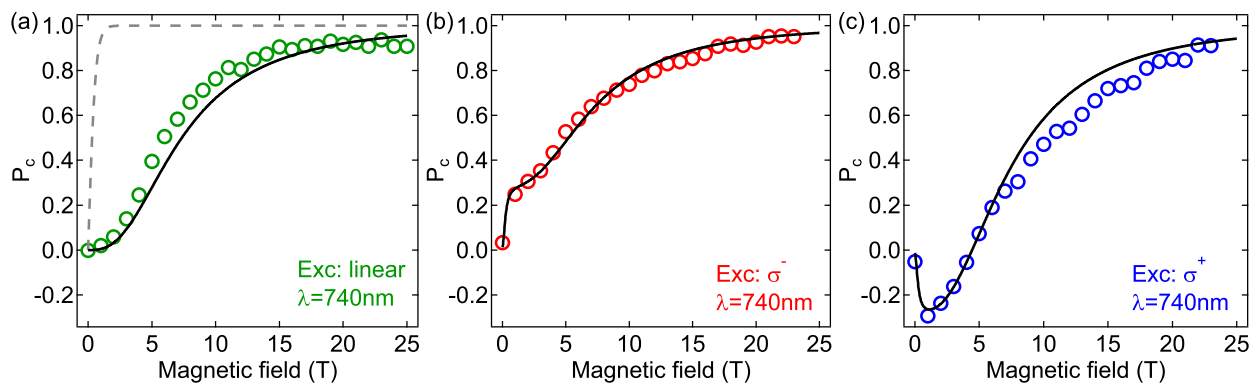}
\caption{Magnetic field dependence of $P_{\text{c}}$ for (a) linearly polarized, (b) $\sigma^-$ and
(c) $\sigma^+$ polarized excitation. The dashed gray line shows the expected evolution of the circular polarization for an
exciton population fully thermalized with the lattice. The measured circular polarization (symbols) is fitted with results of a three
level rate equations model (lines).}\label{fig:Pc}
\end{figure*}

The zero field PL and reflectivity contrast spectra of our structure are presented in the Supporting Information. The PL spectrum consists of sharp peaks attributed to
the recombination of free and charged intralayer excitons in MoS$_2$ and MoSe$_2$ \cite{surrente2017defect}, and a low energy
peak at $\sim\SI{1.38}{\eV}$, which results from the radiative recombination of the interlayer exciton
\cite{baranowski2017probing}. The red shifted interlayer exciton in our heterostructure as compared to a mechanically exfoliated MoS$_2$/MoSe$_2$ bilayer deposited on SiO$_2$ \cite{mouri2017} could be partially attributed to the larger dielectric screening induced by sapphire. This is consistent with the trend observed in TMD monolayers when additional dielectric screening was purposefully introduced \cite{raja2017coulomb}. The magnetoPL is excited  with circularly or linearly polarized light, and is detected using a circular polarization basis. A representative set of magnetoPL spectra of the interlayer exciton is presented in Fig.\
\ref{fig:MagnetoPLenergies}(b). The PL peak exhibits a significant Zeeman splitting and a considerable valley polarization, which
increases with increasing magnetic field and saturates for $B>\SI{20}{\tesla}$. The slightly lower scaling factor for $B=\SI{24}{\tesla}$ results from a small deviation from the saturated valley polarization reached at high fields, due to experimental uncertainties (see, e.g., the valley polarization in Fig.\ \ref{fig:Pc} and the valley polarization extracted from the data of Fig.\ \ref{fig:MagnetoPLenergies}(b) shown in Fig.\ S4(c), Supporting Information). To analyze quantitatively our data, we fit a single Gaussian function to the PL spectra of the interlayer exciton and extract the emission energies, which are plotted in Fig.\ \ref{fig:MagnetoPLenergies}(c) as a
function of the magnetic field. The Zeeman shift of the $\sigma^+$ polarization is larger than that of the $\sigma^-$ polarization. This is a consequence of the diamagnetic effect  -- quadratic in magnetic field -- which blue shifts the exciton energy \cite{stier2018magnetooptics}. The observation of the diamagnetic shift reflects the relatively large electron hole separation and will be subject of a separate study. The energy difference between the two polarizations is shown in Fig.\
\ref{fig:MagnetoPLenergies}(d), where a very large valley Zeeman splitting of $\sim\SI{25}{\milli\eV}$ at the highest magnetic
field is observed. In analogy with the standard analysis for intralayer excitons, we define $\Delta E=E_{\sigma^+}-E_{\sigma^-}=g_{\text{eff}}\si{\micro}_{\text{B}}B$, where $g_{\text{eff}}$ denotes the effective interlayer exciton $g$-factor, $\si{\micro}_{\text{B}}\sim\SI{58}{\micro\eV/\tesla}$ the Bohr magneton, and $B$ the magnetic field. The fitting of the data of Fig.\ \ref{fig:MagnetoPLenergies}(d) gives $g_{\text{eff}}=-13.1\pm0.5$. We estimated $g_{\text{eff}}$ also with the center of mass method \cite{aivazian2015magnetic} (see Supporting Information for a more detailed discussion of the fitting procedures and for the corresponding plots), which yielded $g_{\text{eff}}=-13.4\pm0.5$, identical within experimental error to the value determined by fitting. This very large $g_{\text{eff}}$ has been interpreted as stemming from a non-vanishing valley orbital contribution to the overall magnetic moments of the bands for
heterostructures with a \SI{60}{\degree} stacking angle \cite{nagler2017giant}. This configuration makes transitions between
bands with different valley indexes optically bright, as shown in Fig.\ \ref{fig:MagnetoPLenergies}(a) and discussed more in
detail in the Supporting Information. In our heterostructure, the moir{\'e} pattern yields locally an AB configuration of the registry of the central MoSe$_2$ layer with one of the two MoS$_2$ layers, effectively similar to a lattice matched heterobilayer with \SI{60}{\degree} stacking angle. These spots are expected to be optically bright \cite{yu2017moire,wu2018theory}, and to exhibit a large Zeeman splitting, consistent with the data summarized in Fig.\ \ref{fig:MagnetoPLenergies}. The smaller $g_{\text{eff}}$ observed here, compared with that of a MoSe$_2$/WSe$_2$
heterostructure, is consistent with the smaller difference of the effective mass of the constituents of our sample \cite{ramasubramaniam2012large,kormanyos2015k}.

%%%%%%%%%%%%%%%%%%%%%%%%%%%%%%%%%%%%%%%%%%%%%%%%%%%%%%%%%%%%%%%%%%%%%%%%
%                                                                               %
%                              Lukasz wrote this                               %
%                                                                               %
%%%%%%%%%%%%%%%%%%%%%%%%%%%%%%%%%%%%%%%%%%%%%%%%%%%%%%%%%%%%%%%%%%%%%%%%%%

As seen in Fig.\,\ref{fig:MagnetoPLenergies}(b), applying a magnetic field results in a sizable difference of the PL intensities
of the interlayer exciton recorded in $\sigma^+$ and $\sigma^-$ polarizations. For a quantitative analysis, we define the degree
of circular polarization as $P_{\text{c}} = (I^+-I^-)/(I^++I^-)$, where $I^{\pm}$ denote the PL intensities in $\sigma^{\pm}$
polarizations, respectively. In Fig.\,\ref{fig:Pc}, we plot the magnetic field dependence of $P_{\text{c}}$, for excitation in
resonance with the A-exciton of the MoSe$_2$ monolayer for (a) linear, (b) $\sigma^-$, and (c) $\sigma^+$ excitation
polarization. In the case of intralayer excitons, for an excitation with circularly polarized light, at $B=\SI{0}{\tesla}$ we
expect $P_{\text{c}} \neq 0$, which results from the optical orientation of the valley pseudospin
\cite{mak2012control,cao12,zen12}. This effect has been also observed in heterobilayers \cite{rivera2016valley} and in our
trilayer sample \cite{baranowski2017probing}. In Fig.\,\ref{fig:Pc}(b,c), the observed optically oriented polarization at
$B=\SI{0}{\tesla}$ is very small, but it is recovered by applying a small field and already at \SI{1}{\tesla} $P_{\text{c}}
\approx \pm 0.3$ for $\sigma^{\mp}$ excitation polarizations, respectively. The observed polarization is opposite to
the polarization of the excitation beam, in agreement with our recent report \cite{baranowski2017probing}. This counterpolarized emission might be due to the effects of the moir{\'e} pattern, which only locally preserves the threefold symmetry of the original crystals \cite{yu2017moire,wu2018theory}. A local AB stacking (corresponding to a \SI{60}{\degree} stacking angle and consistent with the observations of Fig.\ \ref{fig:MagnetoPLenergies}) represents a local potential minimum for the interlayer exciton, and it is characterized by a large oscillator strength of the interlayer exciton transition. It also couples to circularly polarized light of opposite helicity with respect to that of the excitation laser \cite{yu2017moire,wu2018theory}. A possible explanation of the opposite polarization is that the optical excitation creates intralayer excitons in the monolayers. The charge carriers are rapidly separated \cite{hong2014ultrafast}, forming interlayer excitons across the heterostructure, which relax to minima of the potential induced by the moir{\'e} pattern. These locations correspond to optically bright spots and couple primarily to light of opposite polarization \cite{yu2017moire,wu2018theory}.

As the field is further increased, $P_{\text{c}}$ increases and at $B >\SI{20}{\tesla}$ reaches $P_{\text{c}} \approx 1$ regardless of excitation
polarization. Nonzero $P_{\text{c}}$ originates from an occupation difference between the interlayer exciton states in different valleys. In
the following, we label these states as pseudospin up $\ket{\uparrow}$ and down $\ket{\downarrow}$, neglecting for the sake of
simplicity the complicated spin/valley structure of the recombining interlayer exciton states depicted in
Fig.\,\ref{fig:MagnetoPLenergies}(a). As the field is increased, the population imbalance increases due to the preferential
occupation of the lower lying Zeeman state (the $\ket{\uparrow}$ state). If thermal equilibrium is established between the
interlayer exciton system and the lattice, $P_{\text{c}}$ is determined solely by the effective $g$-factor of the interlayer
exciton and by the lattice temperature. The dashed line in Fig.\ \ref{fig:Pc}(a) shows the magnetic field dependence of this
polarization, given by $P_{\text{c}}^{\text{eq}} = \tanh(\Delta E/(2\text{k}_{\text{B}} T))$, where $\Delta E$ is the Zeeman splitting
between $\ket{\downarrow}$ and  $\ket{\uparrow}$ states and $T=\SI{4.5}{\K}$ is the bath temperature. The discrepancy between the
experimental results shown in Fig.\,\ref{fig:Pc}(a) and $P_{\text{c}}^{\text{eq}}$ visible at $B \lesssim \SI{15}{\tesla}$
suggests that either the equilibrium is not established or the interlayer exciton system is characterized by a spin temperature
significantly larger than \SI{4.5}{\K}. Even though the interlayer exciton energy (and/or its $g$-factor) suffers from inhomogeneous broadening, we find its impact negligible on the expected equilibrium $P_{\text{c}}$.

\begin{figure}[h!]
\includegraphics[width=0.5\linewidth]{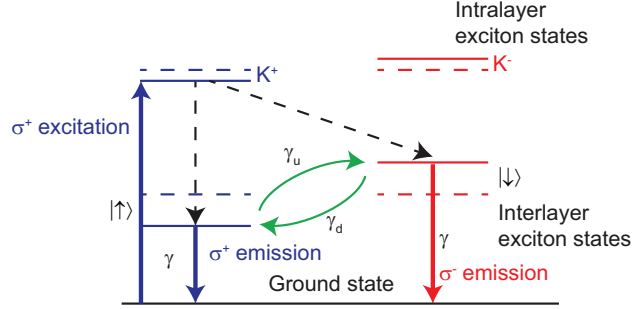}
\caption{Schematic of the rate equations model for the case of $\sigma^+$ excitation. Dashed and
solid horizontal lines denote the relevant exciton states at $B=\SI{0}{\tesla}$ and at $B>\SI{0}{\tesla}$, respectively. A laser beam creates
excitons in K$^+$ valley of one of the constituent monolayers. The type-II band alignment in the heterostructure leads to an
ultrafast charge separation and creation of an interlayer exciton. This process is denoted by dashed arrows. The recombination of the interlayer exciton, denoted by thick, straight arrows, occurs with a rate
$\gamma$ and competes with intervalley scattering, denoted by curved thin arrows, occurring with rates $\gamma_{\text{u}}$ and
$\gamma_{\text{d}}$ for the scattering upward and downward in energy, respectively.}\label{fig:LevelSchematics}
\end{figure}

In order to determine the underlying relaxation processes responsible for the observed field dependence of $P_{\text{c}}$, we
employ a four-level rate equation model \cite{aiv15,neu17} depicted schematically in Fig.\ \ref{fig:LevelSchematics} (see Supporting
Information for further details). The solution of this model provides the field dependence of the interlayer exciton populations
$n_{\uparrow}$ and $n_{\downarrow}$ of the Zeeman split $\ket{\uparrow}$ and  $\ket{\downarrow}$ states. Assuming that
$P_{\text{c}} = (n_{\uparrow}-n_{\downarrow})/(n_{\uparrow}+n_{\downarrow})$, we find \cite{aiv15,neu17}:
\begin{equation}
P_{\text{c}} = P_0 \frac{\gamma}{\gamma+\gamma_{\text{d}}+\gamma_{\text{u}}} + \frac{\gamma_{\text{d}} -
\gamma_{\text{u}}}{\gamma+\gamma_{\text{d}}+\gamma_{\text{u}}},
\label{eq:Pc}
\end{equation}
where $\gamma_{\text{u,d}}$ are intervalley scattering rates (see Fig.\ \ref{fig:LevelSchematics}) and $\gamma$ denotes the recombination
rate. The first term describes the optically created polarization, in which $P_0$ is the polarization transferred from the
circularly polarized excitation. The second term describes the tendency of the system to reach thermal equilibrium. The
intervalley scattering rates can be written as \cite{bla94,wor96}
\begin{equation}
\gamma_{\text{u,d}} = \frac{1}{\tau_{\text{v}0}} \frac{\Gamma^2}{\Gamma^2+\Delta E^2} + \frac{\alpha \Delta
E^3}{\left|\exp\left(\frac{\pm \Delta E}{k_{\text{B}} T}\right)-1\right|}, \label{eq:ts}
\end{equation}
where the first term describes the effect of the electron-hole exchange interaction, which acts as an effective in-plane
magnetic field on the valley pseudospin with a zero-field intervalley relaxation time $\tau_{\text{v}0}$. The precession of the
valley pseudospin around this effective field, together with the reorientation of this field due to momentum scattering
\cite{gla14}, induces the intervalley scattering. However, this is a zero-energy process and thus it is only efficient when the
two valley states $\ket{\uparrow}$ and $\ket{\downarrow}$ are close in energy. Therefore this resonant process is controlled by
the width parameter $\Gamma$. As the Zeeman splitting $\Delta E$ is increased, the exchange-driven relaxation slows down and becomes
negligible when the field-induced splitting $\Delta E$ becomes much larger than  $\Gamma$. The second term in Eq.\ \eqref{eq:ts}
describes the one-phonon spin-lattice relaxation process \cite{orb61}, which requires an emission or an absorption of a phonon,
if the scattering occurs to the lower ($\gamma_{\text{d}}$) or higher ($\gamma_{\text{u}}$) valley, respectively. Consequently,
the scattering rates are proportional to the phonon Bose occupation factors $n_k+1$ or $n_k$, respectively. In Eq.\
\eqref{eq:ts}, $\alpha$ is a measure of the exciton-phonon coupling strength, independent of $\Delta E$.

\begin{figure*}[h!]
\includegraphics[width=1.0\linewidth]{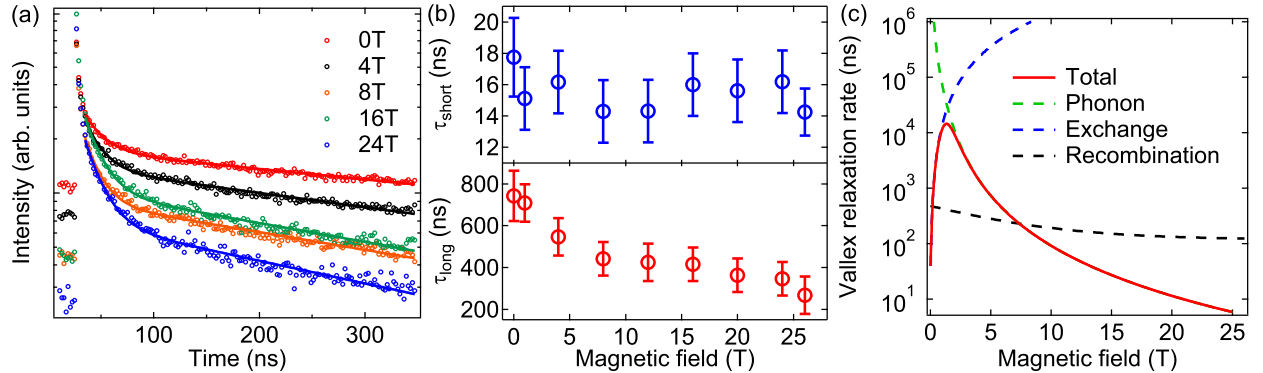}
\caption{(a) Temporal decays of the interlayer exciton PL for various magnetic fields plotted together with fitted biexponential decay functions. (b) Magnetic field dependence of the
short and long decay times. (c) Comparison between the intervalley scattering time $\tau_{\text{v}}$ and the recombination time $\tau_{\text{avg}}$. The contributions to the intervalley scattering
from the exchange-driven process and the spin-lattice relaxation process are also shown.
$\tau_{\text{avg}}$ is interpolated from the measured PL decay times. } \label{fig:times}
\end{figure*}

We evaluate experimentally the recombination rate $\gamma$ by measuring the time-resolved PL of the interlayer exciton. The
temporal decays of the PL signal are shown in Fig.\ \ref{fig:times}(a). The PL decays are fitted well with a biexponential function.
The short decay time obtained from the fitting, $\tau_{\text{short}}$, presented in Fig.\ \ref{fig:times}(b), is about
\SI{16}{\nano\s}, and insensitive to the magnetic field. The fitted long decay time, $\tau_{\text{long}}$, as shown in Fig.\
\ref{fig:times}(b), decreases from about \SI{700}{\nano\s} to \SI{350}{\nano\s} over the investigated field range. In agreement with
previous reports \cite{miller2017long,baranowski2017probing}, both decay times are 3--5 orders of magnitude longer than the decay
of excitonic PL from a monolayer, a consequence of the spatial separation of the electron and the hole. The shortening of
$\tau_{\text{long}}$ with increasing magnetic field can be attributed to the magnetic field induced shrinking of the exciton orbital wave function \cite{aks95}.
Due to a small time window of \SI{350}{\nano\s} available to detect the PL in these experiments, the
evaluation of the long decay time is subject to significant errors.

For the purpose of fitting the field dependence of $P_{\text{c}}$ with our model calculations, we define an average decay time as
$\tau_{\text{avg}} = (A_{\text{long}} \tau_{\text{long}}+A_{\text{short}}
\tau_{\text{short}})/(A_{\text{long}}+A_{\text{short}})$, where $A_{\text{long}}$ and $A_{\text{short}}$ are the amplitudes of
the long and short decays, respectively, and assume that $\gamma = 1/\tau_{\text{avg}}$. To obtain $\gamma$ for all the fields at
which $P_{\text{c}}$ was recorded, we interpolate the field dependence of $\tau_{\text{avg}}$ with an exponential function (see
Supporting Information).

The fitting of the model is performed globally to the three sets of data presented in Fig.\ \ref{fig:Pc}(a-c). The fitting
parameters are $\tau_{\text{v}0}$, $\Gamma$, $\alpha$, and the transferred polarization $P_0 = 0$ for linear excitation and $\mp
P_0$ for $\sigma^+$ and $\sigma^-$ excitations, respectively. The agreement between the fitted curves (solid lines in Fig.\
\ref{fig:Pc}(a-c)) and the experimental data is very good, which allows us to reconstruct the field-induced changes of the
intervalley scattering rates. At fields $\ll \SI{1}{\tesla}$, the optically created valley polarization is quenched by the
exchange-driven process \cite{gla14}. In TMD monolayers, this process leads to intervalley scattering times of the order of a few
ps \cite{zhu2014exciton}. In heterostructures, due to the spatial separation of the electron and the hole, the efficiency of this
process is dramatically reduced. The zero-field intervalley relaxation time obtained from our fitting is $\tau_{\text{v}0} =
\SI{40}{\nano\s}$, four orders of magnitude larger than that of excitons in a WSe$_2$ monolayer \cite{zhu2014exciton} and similar
to the values reported for a WSe$_2$/MoSe$_2$ heterostructure \cite{rivera2016valley}. As the field is increased up to $\sim
\SI{5}{\tesla}$, the exchange-driven mechanism is suppressed and the optically created polarization is recovered. Thus, for
$\sigma^{\pm}$ excitations, $P_{\text{c}}$ becomes negative or positive, respectively, and our fitting yields $P_0 = \mp 0.29$.
The value of $\Gamma$ obtained from the fitting is \SI{40}{\micro\eV}. This value is much smaller than the line width of the
interlayer transition, which suffers from a strong inhomogeneous broadening caused most probably by the laterally varying
distance between the monolayers constituting the heterostructure and resulting from residual adsorbates between them. Also, the
obtained $\Gamma$ is smaller than the exciton homogeneous linewidth measured for monolayer WSe$_2$ \cite{moody2015intrinsic}. We
remark that a similar recovery of the optically created valley polarization was observed for long-lived localized excitons in
monolayer tungsten dichalcogenides \cite{smo17}. The recovery was attributed to a suppressed intervalley scattering of dark
excitons. This mechanism might contribute to the recovery of the polarization in our heterostructure, given that the optically active AB stacking mimics the band structure of W-based TMDs, with dark exciton states lying
energetically below the bright ones \cite{liu2013three}. The recovery of $P_{\text{c}}$ at small fields suggests that
the spin-lattice relaxation is slower than the recombination in that magnetic field range, which leads to a non-equilibrium occupations of the
interlayer exciton states. As the field is increased above $\sim \SI{5}{\tesla}$, the rate of spin-lattice relaxation increases
and drives the system towards equilibrium. Our model assumes that for $\Delta E \gg k_{\text{B}} T$, the intervalley scattering
rate increases as $\Delta E^3$. When this rate becomes larger than $\gamma$, the $P_{\text{c}}$ becomes positive and no longer
determined by the excitation polarization. Our fitting yields $\alpha = \SI{5}{\nano\s^{-1}\eV^{-3}}$. A very similar value was
obtained for excitons in GaAs quantum wells \cite{wor96}, which is a surprising coincidence since we would expect a stronger
exciton-phonon coupling in TMDs \cite{moody2015intrinsic}.

Using Eq.\ \eqref{eq:ts} and our fitted parameters, we calculate the field dependence of the inter-valley scattering time as
$\tau_{\text{v}} = 1/(\gamma_{\text{d}}+\gamma_{\text{u}})$ and compare it to the interpolated average recombination time $\tau_{\text{avg}}$
in Fig.\ \ref{fig:times}(c). As inferred from the analysis of the evolution of $P_{\text{c}}$ with magnetic field, $\tau_{\text{v}}
< \tau_{\text{avg}}$ at $B<\SI{1}{\tesla}$, where the exchange-driven process dominates. Then, in the intermediate field range
$\SI{1}{\tesla} \leq B \lesssim \SI{7}{\tesla}$, $\tau_{\text{v}}$ becomes larger that $\tau_{\text{avg}}$. Above $\sim
\SI{7}{\tesla}$, the intervalley scattering time decreases below the recombination time due to the increased Zeeman splitting
$\Delta E$, and the interlayer exciton system is driven toward thermal equilibrium, as evidenced by the experimental data which approach
the dashed curve in Fig.\ \ref{fig:Pc}.

In conclusion, we have performed detailed magnetoPL spectroscopy of interlayer excitons formed in a MoS$_2$/MoSe$_2$/MoS$_2$
heterostructure. The effective $g$-factor of the interlayer exciton  is significantly larger as compared to that of intralayer
excitonic transitions, due to a non-vanishing contribution of the valley orbital moments. The large electron-hole separation in the interlayer exciton results in the
suppression of intervalley scattering, which is directly reflected in our polarization resolved measurements. At zero magnetic
field, the polarization degree is mainly determined by the intervalley scattering driven by the electron-hole exchange interaction. A small magnetic field can
efficiently suppress this mechanism, which results in a recovery of the optically induced polarization. In the high field limit,
the PL polarization is dominated by the thermal occupation of Zeeman split states and does not depend on the injected
polarization. The interplay between exchange interaction, phonon driven scattering and magnetic field can result in non-monotonic
behavior of the valley polarization. Our results represent a crucial step for the implementation of valleytronic concepts based
on interlayer excitons by providing a thorough insight into the valley polarization generated via an applied magnetic field and
the competing depolarization mechanisms.

\section*{Methods}\label{sec:methods}
\textit{Sample preparation.} The large area MoS$_2$ and MoSe$_2$ monolayer films, which form the investigated heterostructure, were grown
separately by chemical vapor deposition (CVD) on sapphire substrates \cite{dumcenco2015large,mitioglu2016magnetoexcitons}. The
heterostructure was fabricated by two sequential KOH-based transfer steps \cite{wang2016interlayer}. The first transfer allowed
to form a MoS$_2$/MoSe$_2$ heterobilayer, which was subsequently transferred onto a MoS$_2$ film to form the trilayer stack \cite{surrente2017defect}. This approach allowed us to obtain multiple van der Waals heterostructures with varying stacking angles over very large areas \cite{surrente2017defect,baranowski2017probing}.

\textit{Magneto-optical spectroscopy.} The sapphire substrate with large area heterostructures was mounted on a $x-y-z$
translation stage driven by piezoelectric actuators and cooled down at \SI{4.5}{\K} in a liquid helium cryostat. Static magnetic
fields up to \SI{28}{\tesla} were applied making use of a water cooled resistive magnet. The PL was excited by a diode laser emitting
at \SI{640}{\nano\m} (excitation in resonance with the MoS$_2$ A-exciton) or a Ti:sapphire laser tuned at \SI{740}{\nano\m}
(excitation in resonance with the MoSe$_2$ A-exciton). The typical excitation power was kept as low as possible and in any case lower than \SI{30}{\micro\W}. The excitation laser was circularly or linearly polarized with a linear
polarizer and a Babinet-Soleil compensator. The beam was focused on the sample by a long working distance microscope objective,
which was also used to collect the PL. This signal was analyzed in a circular polarization basis with a zero-order
quarter-waveplate and a linear polarizer and dispersed by a \SI{0.3}{\m} long monochromator coupled to a liquid nitrogen cooled
CCD detector. Time resolved magnetoPL was measured by operating the diode laser in pulsed mode with a repetition frequency of
\SI{2.5}{\MHz}, synchronized with a Si avalanche photodiode. The PL was spectrally selected by making use of a longpass filter.

\begin{suppinfo}
Low temperature \si{\micro}PL spectra of a Mo$_2$/MoSe$_2$/MoS$_2$ heterostructure at zero magnetic field, estimation of valley Zeeman splitting for intralayer and interlayer excitons, additional details concerning the fitting procedures used for the determination of the PL energy, four level rate equation model and magnetic field dependence of the valley polarization for an excitation in resonance with the A exciton of MoS$_2$.
\end{suppinfo}

\begin{acknowledgement}
This work was partially supported by the R{\'e}gion Midi-Pyr{\'e}n{\'e}es under contract MESR 13053031, BLAPHENE, and STRABOT
projects, which received funding from the IDEX Toulouse, Emergence program, by ``Programme Investissements d'Avenir'' under the
program ANR-11-IDEX-0002-02, reference ANR-10-LABX-0037-NEXT and by the PAN--CNRS collaboration within the PICS 2016-2018
agreement. N.Z.\ holds a fellowship from the Chinese Scholarship Council (CSC). This work was also financially supported by the
Swiss SNF Sinergia Grant 147607. M.B.\ acknowledges support from Polish Ministry of Higher Education and Science through grant
DEC 1648/MOB/V/2017/0. This work was also supported by HFML-RU/FOM and LNCMI-CNRS, members of the European Magnetic Field
Laboratory (EMFL).
\end{acknowledgement}

\section{Conflict of interest disclosure}
The authors declare no competing financial interest.

%\bibliography{BibMagnetoPL_IE}
\providecommand{\latin}[1]{#1} \makeatletter \providecommand{\doi}
  {\begingroup\let\do\@makeother\dospecials
  \catcode`\{=1 \catcode`\}=2\doi@aux}
\providecommand{\doi@aux}[1]{\endgroup\texttt{#1}} \makeatother
\providecommand*\mcitethebibliography{\thebibliography} \csname
@ifundefined\endcsname{endmcitethebibliography}
  {\let\endmcitethebibliography\endthebibliography}{}

\begin{tocentry}

\includegraphics{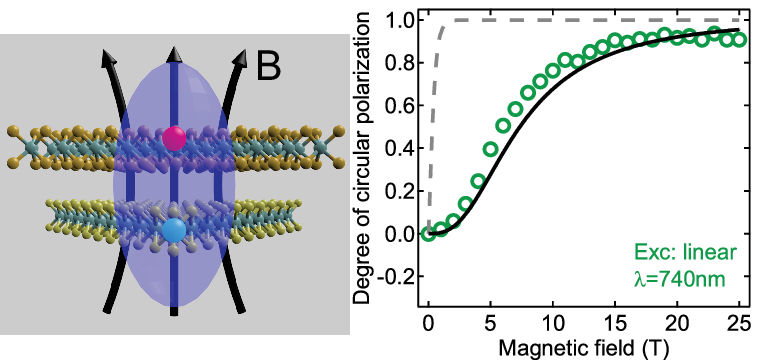}

\end{tocentry}

\end{document}